\documentclass[aps,prl,reprint,groupedaddress]{revtex4-1}
\usepackage{graphicx, gensymb}

\begin{document}

\title{A re-analysis of x-ray heated aluminum scattering data}

\author{Eliseo Gamboa}
\email[eliseo@umich.edu]
\noaffiliation

\date{\today}

\begin{abstract}
A re-anaylsis of the x-ray scattering data in Sperling et al \cite{sperling2015free} indicates that the measurement of elevated temperatures may be unreliable. The original analysis did not include uncertainties in the instrument function used to extract the plasmon signal. Including these effects leads to poor agreement with the theoretical models. The experimental conditions were also not measured and the heating was likely significantly lower than claimed.  
\end{abstract}

\maketitle
  
In a recent paper Sperling et al \cite{sperling2015free} claim to have measured the temperature of aluminum isochorically heated and probed by the LCLS free electron laser (FEL). By focusing the FEL to a small spot, matter can be significantly heated via photoabsorbtion while the x-ray scattering spectrum is measured. The temperature can be inferred from observing the intensity ratio between the upshifted and downshifted x-ray plasmon peaks via detailed balance. 

I will show that the claimed measurement of the upshifted plasmon was likely not possible given the precision of the measurements in the experiment. By neglecting the uncertainties in the spectral measurements, the authors interpreted random and systematic errors for an upshifted plasmon. Including these sources of error by resampling the data, the theoretical models fail to achieve statistical significance. In addition, the best focus and defocussed data do not show a systematic relationship to the 6 eV and 0.3 eV theoretical models, respectively. Two of the defocussed runs, where the temperature should not be significantly raised, show evidence of an upshifted plasmon that is similar to the best focus runs.   

It highly likely that the heating from the focused x-ray irradiation was significantly lower than the reported value of 6 eV. The authors  did not measure the stated x-ray spot size of 1 $\mu$m and used incorrect values for the x-ray energy fluence to initialize the SCFLY simulations. The FEL lenses were placed at the geometric best focus position. The x-ray focal position is known to vary by a few cm from the geometric position, introducing a factor of $\sim$25 on the intensity in the experiment.

\section{Error analysis of the x-ray scattering spectra}

For collective x-ray scattering, the electron temperature can be calculated by measuring the intensity ratio of the upshifted to downshifted x-ray plasmon peaks using the detailed balance relation
\begin{eqnarray}
\frac{S^+(k)}{S^-(k)} = e^{-\hbar \omega_p/k_B T}
\label{detailed_balance_eq}
\end{eqnarray}
where $S^+(k)$ and $S^-(k)$ are the frequency-integrated intensities in the upshifted and downshifted plasmons, respectively, and $\hbar \omega_p \sim$  18 eV is the plasmon energy shift. The signal in each plasmon must be distinguished from the generally stronger elastic peak, found centered at zero energy shift relative to the x-ray probe. The elastic peak is assumed to reflect the spectral distribution of the seeded FEL beam convolved with the resolution of the spectrometer. It is usually referred to as an instrument function.

\begin{figure}[ht!]
	\includegraphics[width=0.4\textwidth]{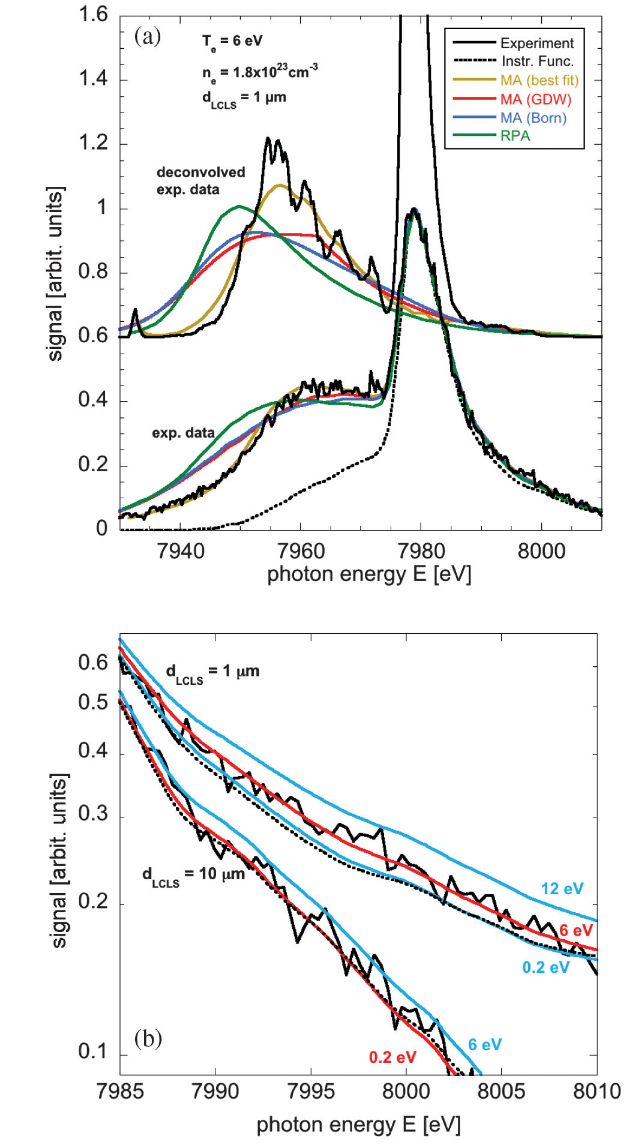} 
\caption{Forward scattering data from aluminum as is shown as Figure 2 of Sperling et al \cite{sperling2015free}. While the downshifted plasmon is evident in the region around 7960 eV in (a), the upshifted plasmon, which would be centered around 8000 eV, is a much more subtle effect. The fit of the upshifted plasmon in (b) uses an over-smoothed instrument function and neglects a consideration of the measurement uncertainties. }
\label{fig2_prl}
\end{figure}

I show the main experimental result of Sperling et al in Figure \ref{fig2_prl}. In (a), the downshifted plasmon centered at 7960 eV is readily apparent. At a best focus FEL spot size of 2 $\mu$m FWHM, a weak upshifted plasmon was claimed to have been observed. This manifested as a small additional signal of the scattering data over the instrument function in the spectral region around 8000 eV. This is shown in detail in Fig. \ref{fig2_prl}(b). 

\begin{figure}[ht!]
	\includegraphics[width=0.5\textwidth]{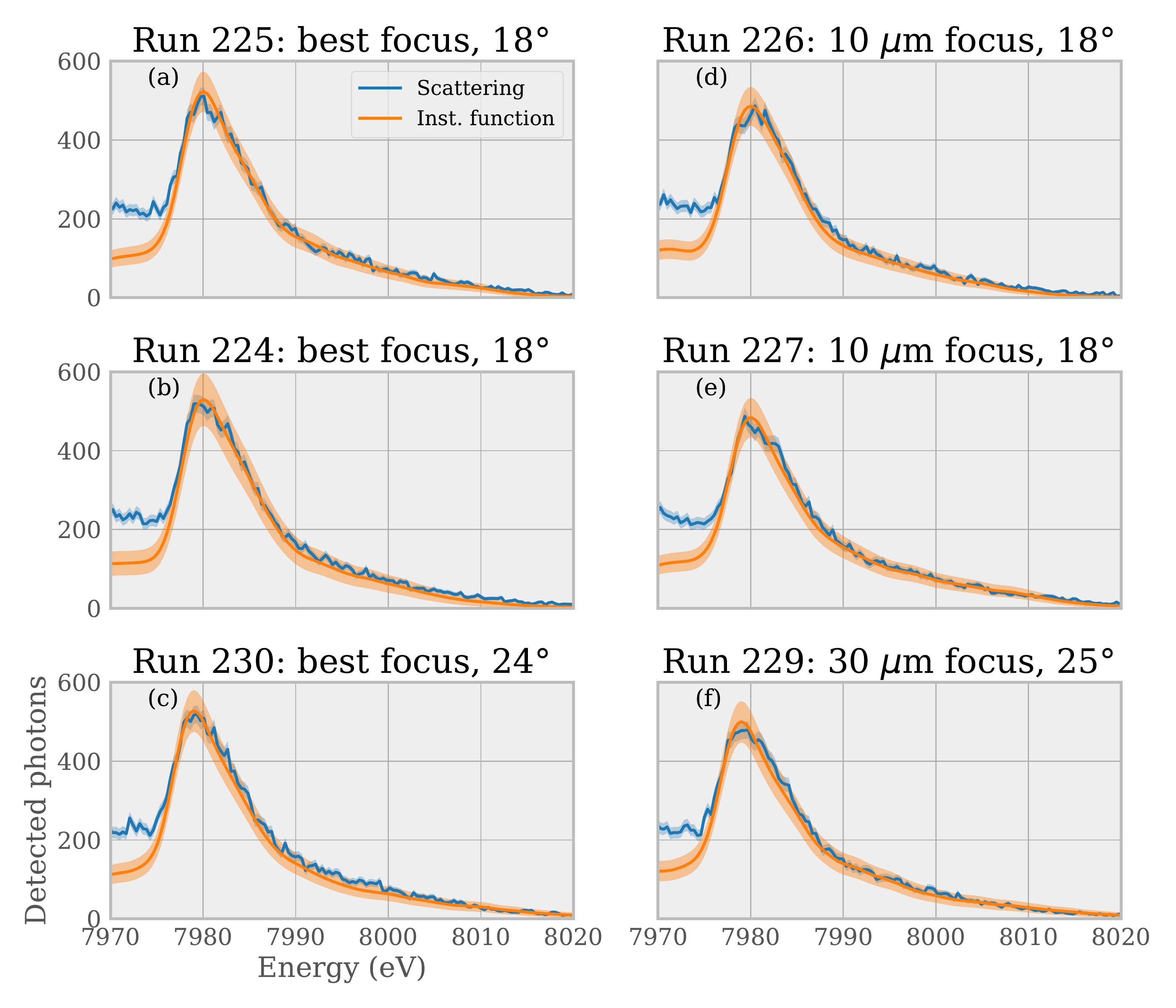} 
\caption{Forward scattering data (blue) plotted against the reconstructed instrument function (orange). Error bars for each curve are shown as the shaded regions. The data shown in (a-c) were collected with the FEL at the nominal best focus position. For comparison, (d-f) were collected with the FEL significantly defocused. Over the spectral region containing a possible upshifted plasmon (8000-8020 eV) the uncertainties in the instrument function are larger than the differences between the two curves.}
\label{fig1}
\end{figure}

The seeded FEL beam is known to have significant intensity and spectral fluctuations \cite{amann2012demonstration}, both in the central peak and the pedestals located in the spectral region also containing the plasmons. The instrument function must be independently characterized for each experimental spectra that is collected. In the present experiment, the FEL spectrum was not directly measured. Instead, a complimentary GaAs spectrometer measured the non-collective x-ray spectrum at a scattering angle of 60\degree \cite{zastrau2014bent}. The FEL spectrum was assumed to be equivalent to the elastic peak measured by the GaAs spectrometer. This elastic peak was isolated by subtracting away the broad Compton peak, which was assumed to be parabolic in shape. 

The instrument function was reconstructed by convolving the extracted elastic peak with the resolution function of the HAPG spectrometer used to measure the forward scattering spectrum. Small systematic errors, for example in the baseline of the non-collective elastic peak or the intensity of the Compton peak, can potentially have large impacts on the derived instrument function. In addition, the resolution function of the HAPG spectrometer was not accurately measured. It was instead extracted by deconvolving a measurement of the Cu K$_{\alpha}$ spectrum from a theoretical model. This reconstruction process for the instrument function was not independently validated to produce the correct result. There is a high likelihood that it introduced systematic errors to the analysis, which are especially concerning given the precision needed to measure the weak upshifted plasmon signal.  

%However, without an independent measurement of the source, the accuracy of the reconstruction process cannot be established. 

The GaAs spectrometer measured the elastic peak with about ten times fewer photons than the HAPG spectrometer. The resolution of the HAPG crystal was about 10 times less than the GaAs crystal. The convolution step over-smooths the natural shot noise in the instrument function. The smoothed instrument function, plotted as a dashed line in Fig. \ref{fig2_prl}, seems to suggest that the instrument function is known more precisely than it is. The error bars can be recovered by summing the absolute photon signal measured by the CCD detector on the GaAs spectrometer. The reconstructed instrument function can then be normalized to this level. The photon signal $N_{ph}$ is then calculated for each spectral bin and the associated error bars of $\sqrt{N_{ph}}$. 

In Figure \ref{fig1} I show scattering data in the spectral region which would contain any upshifted plasmon. Fig. \ref{fig1} (a-c) is all the heated data collected at the nominal best focus; plot (c) is the data shown presented in Fig. 2 in Sperling et al that I reproduced as Fig. \ref{fig2_prl}. For comparison I also show data taken with the FEL defocused (d-f), where the aluminum should be unheated. In each of these cases, it can be seen that the random uncertainties in the instrument function are typically larger than the variations between the scattering and instrument function. 

Since the instrument function was measured with fewer total photons, it has to be scaled up by an arbitrary factor of 5-8 to match the intensity of the elastic peak of the forward scattering spectrum. The uncertainties must also be scaled by the same factor, yielding larger error bars on the instrument function relative to the scattering. I caution that, even drawn with the appropriate error estimates, the solid blue line should not be interpreted as the true value of the instrument function. This curve is still inappropriately smoothed and should not be used to extract plasmon intensity ratio.   

\subsection{Resampling analysis}
%I used a Monte Carlo approach to include these uncertainties in extracting the temperature from the experimental data.
I included the uncertainties in the analysis by resampling and fitting the spectral data. The forward scattering spectrum and instrument function were modeled as ensembles of Gaussian-distributed variables, with means and standard deviations equal to the measured values in each spectral bin. The two curves were resampled from their associated probability distributions and scaled to overlap the mean values in the elastic peaks. I then subtracted the instrument function from the scattering to isolate the plasmon signal. 

As can be seen in Fig. \ref{fig1}, the uncertainties in the peak intensity of the instrument function are large compared to the differences between the scattering and instrument function. The resampling causes the peak intensity of the instrument function to fluctuate. Since the instrument function is scaled to the scattering data, the relative difference between the scattering and instrument function in the upshifted region will also change. Therefore, scaling the instrument function introduces another random source of error to the fitting process that was not included in the original analysis. By averaging over many repetitions, we can include this variation in the fitting.

\begin{figure}[ht!]
	\includegraphics[width=0.5\textwidth]{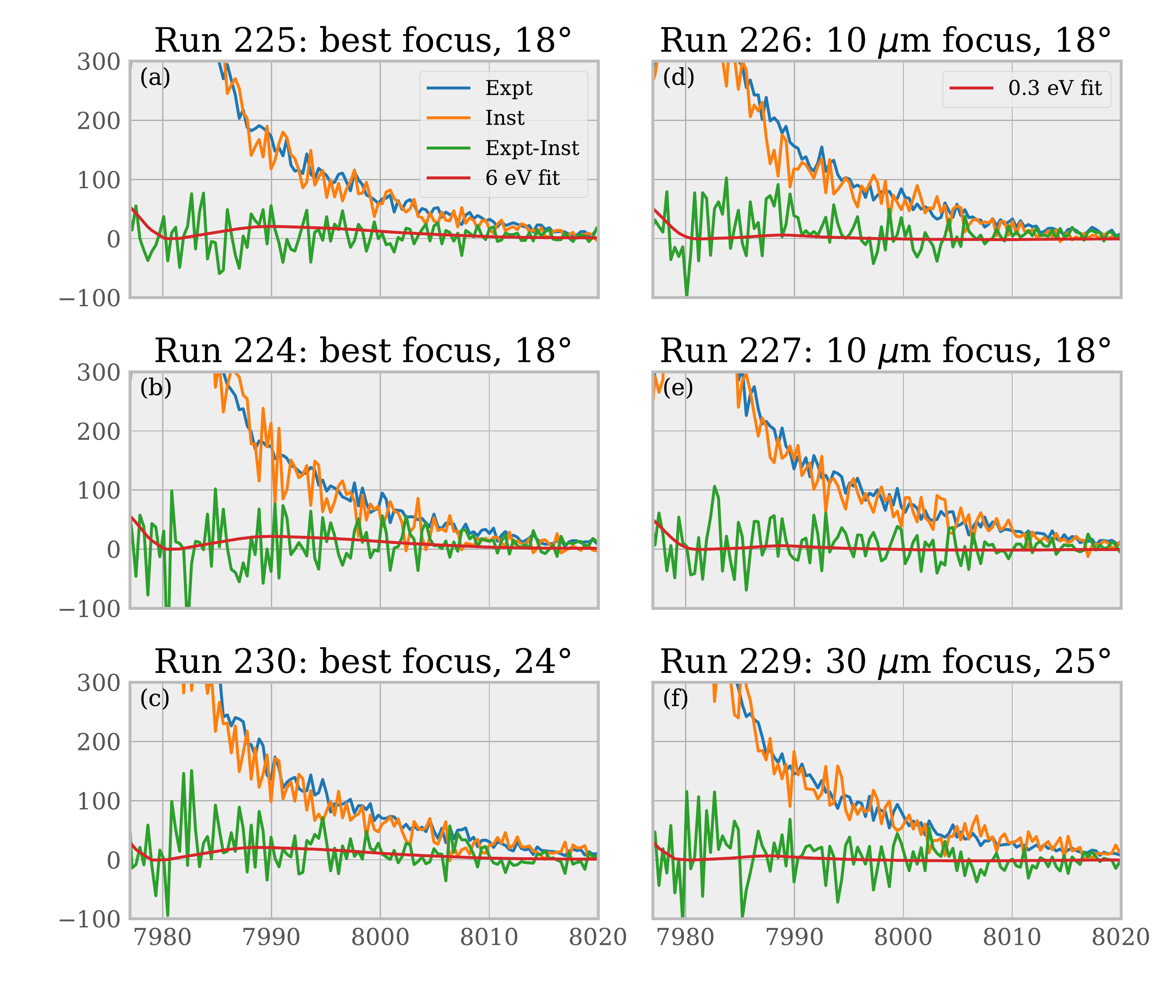} 
\caption{Plotted are the resampled scattering spectra (blue), instrument function (orange), and the difference between the two measured spectra (green). Any evidence of an upshifted plasmon would appear in the subtracted spectrum. The subtracted data are compared to the theoretical fits to the upshifted plasmon (red) from Sperling et al calculated for T = 6 eV (a-c) and 0.3 eV (d-f).}
\label{fig2}
\end{figure}

I plot the resampled data in Figure \ref{fig2}. For the best focus (a-c) and defocused data (d-f) alike, it is difficult to spot by eye any significant differences between the resampled instrument function (orange) and the scattering data (blue). The situation is not much improved when looking at the isolated upshifted plasmon signal (green). The spectra are dominated by random noise, primarily from the poorly constrained instrument function. The plasmon signal does not show good qualitative agreement with the model fits in either case. 

I repeated the resampling procedure 100,000 times. For each trial, I evaluated the goodness-of-fit of both the 6 eV and 0.3 eV theoretical models. These models are identical to those presented in Sperling et al and Fig. \ref{fig2_prl}. By Eq. \ref{detailed_balance_eq}, the intensity of the upshifted plasmon at 0.3 eV is indistinguishable from room temperature.  I used a $\chi^2$ test with N = 100, containing the spectral window in Fig. \ref{fig2}, and $\nu$ = 97. The fitting parameters to the model are the temperature, the theoretical elastic to inelastic scattering ratio, and the scaling between the instrument function and scattering. For p=0.05, $\chi^2 = 121$ is the critical value. The results are summarized in Table \ref{chi_table}

\begin{table}
\centering
\begin{tabular}{ l | c | c | c| r  }
  \hline			
  Run & Focus & $<\chi^2_{\text{0.3 eV}}>$ & $<\chi^2_{\text{6 eV}}>$ & $<p>$  \\
  \hline
  225 & Best & 158 & \textbf{144} & 0.0014 \\
  224 & Best & 171 & \textbf{146} & 0.001 \\
  230 & Best & 146 & \textbf{123} & 0.048 \\
  226 & 10 $\mu$m & 173 & \textbf{144} & 0.0014\\
  227 & 10 $\mu$m & \textbf{133}& 154 & 0.0086\\
  229 & 30 $\mu$m & \textbf{125} & 127 & 0.042 
\end{tabular}
\caption{Summary of the fits to the forward scattering data.  $<\chi^2_{\text{0.3 eV}}>$ and $<\chi^2_{\text{6 eV}}>$ denote the denote $\chi^2$ averaged over the repetitions for the 0.3 eV and 6 eV models, respectively. The smallest value of $\chi^2$ is shown in bold. The corresponding average p-value, $<p>$ is shown for the best fit.}
  \label{chi_table}
\end{table}
 
The theoretical fits all fail to reach a significance level of p$>$0.05 on average. For the best focus data, the best case was run 230 where the 6 eV model gives $\chi^2 = 123$ and p=0.04. The two other best focus runs (224,225) have much worse agreement to the 6 eV model, with $\chi^2 \sim$ 145. The defocused run 229 has a similar agreement to both the 6 eV ($\chi^2$ = 127) and 0.3 eV ($\chi^2 = 125$) models. While this run slightly favors the low-temperature model, it fits the high temperature model better than runs 224 and 2445 and is quite close to run 230. Additionally, the defocused data in run 226 has better agreement to the 6 eV model than the 0.3 eV model. Thus it is doubtful whether we can even establish if there is a difference in the upshifted plasmon region between the best focus and defocussed data sets. Variations between the scattering data and instrument function may only reflect the inaccuracies in the reconstruction of the instrument function itself and noise in the extracted inelastic scattering.  

In contrast to the report of Sperling et al of an elevated temperature at best focus, the data in the upshifted region does not show a good agreement to the 6 eV theoretical model. The fits fail to reach significance and the best focus and defocused data do not show a systematic relationship to the models. The large variations in the spectral data overwhelm the sensitivity of the temperature measurement.

An alternative explanation is that the data are not consistent with an elevated temperature in any case.  As I show next, it is highly likely that the intensity at best focus was significantly less than what was described. The data fail to show a meaningful upshifted plasmon signal because the aluminum was not significantly heated.

\section{Experimental conditions} 

Successfully heating the aluminum sample with the FEL depends on focusing the x-ray to a small spot. The spot size was not measured in the experiment, leaving the actual experimental conditions largely unknown. The SCFLY simulations used to justify a temperature of 6 eV overestimated the x-ray fluence on target by at least a factor of 3.6. Assuming the temperature scales linearly to the fluence, the upshifted plasmon signal at best focus would be weaker by a factor of 5000. The aluminum was likely not heated to a point where an upshifted plasmon could have been detected over the noise in the measurements.  

The experiment used a set of beryllium compound refractive lenses, located approximately 4 m away from the interaction region, to focus the x-ray beam from a diameter of 800 $\mu$m to a nominal 1 $\mu$m. At the time of this experiment, the set of prefocusing lenses now located 100m upstream of the target chamber \cite{heimann2016compound} had not been implemented at the LCLS. The x-ray beam overfilled the 800 $\mu$m aperture of the focusing lenses. The divergence was then about 2 $\mu$m FWHM/cm of defocusing. The best focus is about 1 $\mu$m FWHM and is typically located a few cm away from the calculated best-focus position \cite{sikorski2015focus}. This is a consequence of form errors in the beryllium lenses and positioning errors in the lens holders. 

\begin{figure}
	\includegraphics[width=0.5\textwidth]{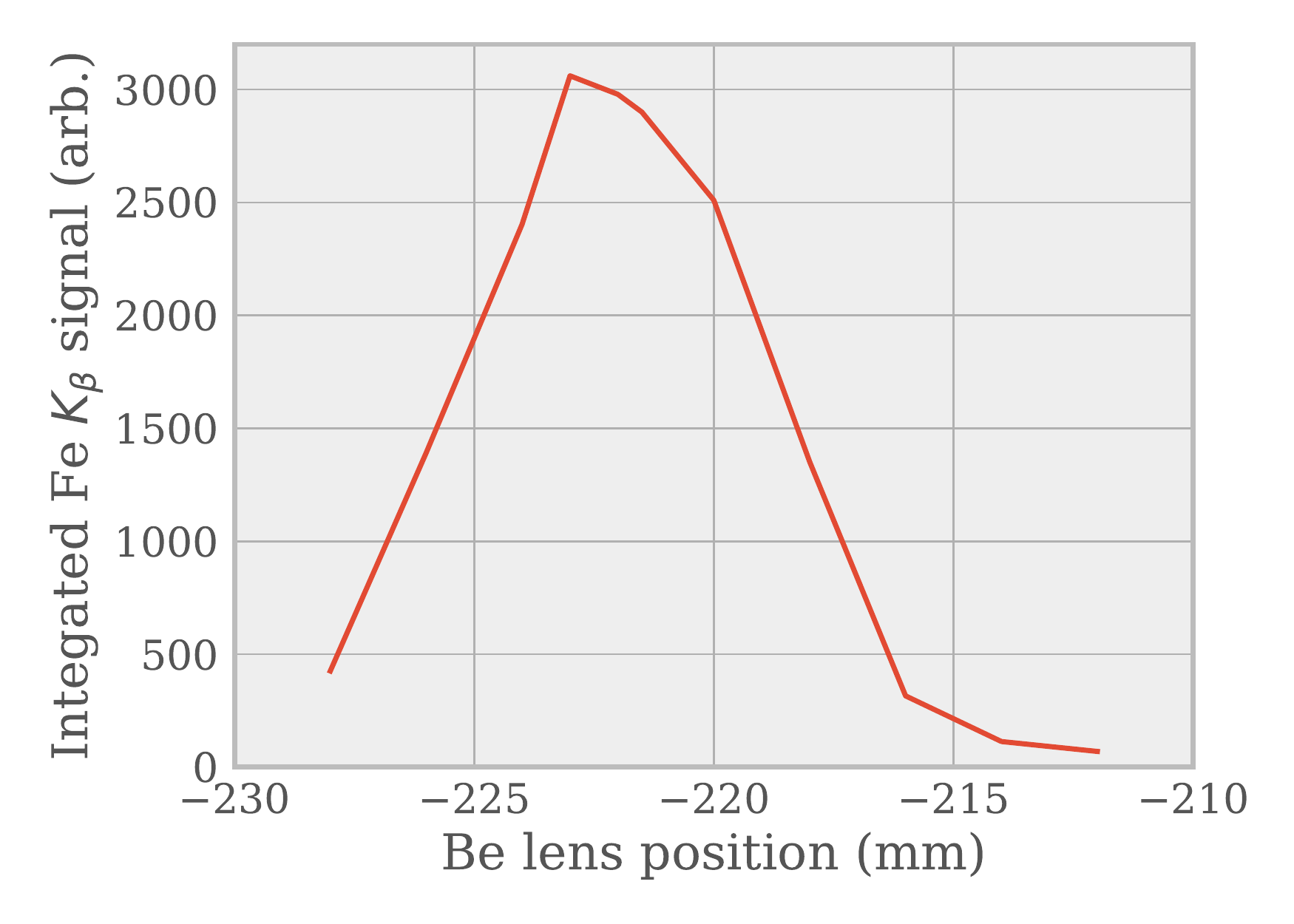} 
\caption{Results from measuring the FEL spot at 7 keV using two-photon $K_{\beta}$ excitation. The experimental best focus at z=-223 maximized production of $K_{\beta}$ x-rays. The calculated best focus position was at z=-205.5 mm.}
\label{fig3}
\end{figure}

The FEL spot size can be optimized in situ by tuning the FEL energy below the K-edge of a transition metal and maximizing the two-photon excitation of the K$_\beta$ transition. In this process, photo-absorption of an x-ray creates a long-lived L-shell vacancy in an atom. This excited atom has a large cross section (~Mb/atom) to absorb a subsequent x-ray, promoting a K-shell electron to fill the L-shell vacancy. A characteristic K$_\beta$ x-ray is then emitted from the de-excitation of the atom. This process scales as the square of the x-ray intensity. 

This technique was used to optimize the x-ray focus in another recent experiment. Figure \ref{fig3} shows the results for scanning the x-ray focus position. An iron foil was irradiated with the FEL at a fixed energy of 7.05 keV. The x-ray production is maximized with the lenses at around z=-222 mm. The experimentally determined best focus was then characterized by examinig the size of the craters ablated by the x-ray beam \cite{heimann2016compound}, yielding a spot size of 2.33 $\mu$m FWHM. In comparison, the calculated best focus for this configuration was estimated to be 0.4 $\mu$m FWHM at z=-205.5 mm, 16.5 mm away from the experimentally determined best focus.  

For the experiment in Sperling et al, the lenses were set at the calculated best-focus position, but the actual dimensions of the focal spot were not measured. The paper asserts that the x-ray spot size was 2 $\mu$m FWHM. Using the geometrical beam divergence, the x-ray spot size in the experiment should be quoted as 1 $+$4/-0 $\mu$m FWHM. Without experimentally optimizing the focus, the x-ray intensity can be then expected to vary by a factor of $5^2$ relative to the optimum position. 
 
The SCFLY simulations used a value of 0.09 mJ for the x-ray energy per pulse incident on the target, or a per pulse energy of 0.3 mJ and a transmission to the target chamber of 30\%. The per-pulse energy was measured in the experiment to be 0.25 mJ. Results presented by Heimann et al indicate the transmission is 20\% \cite{heimann2016compound}, or an effective 0.05 mJ/pulse delivered to the target. Thus the energy input to the simulations was overestimated by about 40\%. 

SCFLY is 0D and therefore does not take into account the variation in the x-ray dose from the thickness of the foil. X-rays near the back side of the foil, where the FEL intensity has been diminished by photo-attenuation, are more heavily weighted in the measured forward scattering spectrum. For a foil of thickness $d$, atoms at a depth $x$ are irradiated with an intensity
\begin{eqnarray}
I(x) = I_0(x) e^{-\mu x}
\end{eqnarray}
For some scattering angle $\theta$, the contribution to the scattering spectrum from a thickness $dx$ has to be weighted by the sample transmission
\begin{equation}
S(x) dx = I_0 e^{-\mu x} e^{-mu (d-x)/\cos\theta} dx
\end{equation}
The depth-averaged dose is then 
\begin{eqnarray}
S_{ave} = 1/d \int_0^d S(x) dx
\end{eqnarray}
Using the parameters from the experiment (d = 50 $\mu$m, $\mu$ = 132 cm$^-1$, $\theta$ = 24 $\deg$), depth averaging reduces the effect x-ray dose by another 50\%. Considering the corrected pulse energy and depth averaging, the net x-ray fluence is therefore reduced by a factor of 3.6. 

Finally, The SCFLY calculations indicate a maximum temperature of 6 eV by the end of the pulse, which was claimed to be consistent with an upshifted plasmon and an experimental temperature of 6 eV. The experimental measurements are time-integrated over the duration of the FEL pulse. Measuring an average temperature of 6 eV is inconsistent with simulations suggesting a final temperature of 6 eV.  From the detailed balance relation, the time-averaged intensity ratio is 
\begin{eqnarray}
\frac{S(k)^+}{S(k)^-} = \frac{\int_0^\infty  e^{-\hbar \omega_p/k_B T(t)} FEL(t) dt }{\int_0^\infty FEL(t) dt}
\end{eqnarray}
where $FEL(t)$ and $T(t)$ are the FEL intensity and temperature temporal profiles. For a final temperature of 6 eV, the time-averaged SCFLY simulations would be consistent with an experimental intensity ratio of 0.01.  

An intensity ratio of 0.05 was claimed to be observed in the data. This would require a final temperature of 10 eV, or an even tighter x-ray focus than is likely possible. 

\subsection{Prevalence of successful x-ray heating}
Given that the actual x-ray focal spot was not characterized, there is a large degree of uncertainty whether the chosen best focus position was small enough to create elevated temperatures in the aluminum sample. Assuming that the actual focal position is a Gaussian distributed variable centered at the calculated position, the probability that the focal spot had a FWHM of at most $d$ is
\begin{eqnarray}
P(d) = \sqrt{2\pi}\sigma\int_{-\ell}^\ell e^{-x^2/2*\sigma^2} dx \\
\ell = (d-d')
\end{eqnarray}
Here $\sigma$ = 4 $\mu$ and $d'$ is the best focus spot of 1 $\mu$m. 

The probability that the focal spot was at most 2 $\mu$m FWHM in size is 20\%. To compensate for the overestimate in the x-ray energy on the target for the SCFLY simulations, the focal spot would have to be at most 1.2 $\mu$m FWHM. The probability for this to have occurred is  4\%.

\section{Conclusion}

The temperature measurement of Sperling et al are not supported by the data. The quality of the data were too poor to perform a meaningful measurement of the upshifted plasmon at the claimed conditions.  As the spot size was not measured during the experiment, it is unknown what the actual conditions were at the position of the calculated best focus. While it is possible that the x-ray spot size was small enough for an appreciable level of heating, the known errors in the focus make this very unlikely. Due to the complexity of the experimental setup and subsequent changes to the experimental facilities, subsequent measurements of the x-ray spot size may not be directly applicable. The theoretical work that was presented may be sound, but should not be considered to be validated by the experimental data.

\begin{acknowledgments}
The author wishes to acknowledge the assistance of LCG, HON, ADB, WF, and SBH. This work is dedicated to the memory of Mikhail V Konnik.
\end{acknowledgments}


\bibliographystyle{unsrt}
%\bibliography{bibfile}

\end{document}